# PEDOT:PSS-Coated Magnetoelastic Sensors for Highly Sensitive Wireless Humidity Sensing

Wenderson R. F. Silva[1*], Robson C. O. Guedes[1], Gilberto Rodrigues-Junior[1], Eduarda P. M. Campos[2], Ângelo Malachias[2], Joaquim B. S. Mendes[1*]

[1]Departamento de Física, Universidade Federal de Viçosa, 36570-90, Viçosa, Minas Gerais, Brazil

[2]Departamento de Física, Universidade Federal de Minas Gerais, 31270-901 Belo Horizonte, Minas Gerais, Brazil

## ABSTRACT

Magnetoelastic (ME) resonators enable passive and wireless sensing, making them attractive platforms for monitoring relative humidity (RH) in sealed or hard-to-access environments. Here, we report ME humidity sensors functionalized with the polymer poly(3,4-ethylenedioxythiophene):polystyrene sulfonate (PEDOT:PSS), used for the first time in this class of ME devices. PEDOT:PSS films were deposited on Metglas resonators by a simple drop-casting method, yielding uniform and stable coatings. Structural and surface analyses (SEM, EDX, Raman, AFM/KPFM, and XRD) confirmed the integrity of the polymer and revealed humidity-driven swelling of PSS domains, accompanied by increased chain mobility and lamellar spacing (from 23.5 Å to 24.2 Å at 95% RH). Resonance measurements demonstrated a clear frequency downshift with increasing RH, governed by combined mass loading and viscoelastic damping effects. The optimized sensor exhibited a sensitivity of 155 Hz/% RH in the 20 - 70% RH range, surpassing values reported in ME humidity sensors, with a resolution below 0.1% RH and excellent response and recovery times. These findings establish PEDOT:PSS as an effective functional layer for ME-based humidity sensors, enabling low-cost, wireless, and highly sensitive devices for practical monitoring applications.

*Corresponding authors: wenderson.f@ufv.br , joaquim.mendes@ufv.br

## 1. Introduction

Precise monitoring of RH is critical for productivity and safety in sectors such as healthcare, agriculture, industrial processing, and food storage. To meet this growing demand, a wide variety of sensors have been developed, exploiting physical phenomena such as electric resistivity, capacitance, FET properties, and oscillator circuits [1–4]. Beyond traditional polymeric materials, sensors have been fabricated from rGO/ZnO [5,6], Ag/$SnO_2$ nanocomposites [7], GaN nanorods [8], two-dimensional boron [9], triboelectric chitin nanopaper [10], and biocompatible devices derived from natural potato peel [11]. However, most of these technologies still rely on physical connections for power and data transfer, which severely limits their implementation in hermetic or sealed environments such as food packaging or controlled-atmosphere storage systems [12]. This limitation has become a strong motivation for the development of novel wireless and passive ME sensing platforms capable of operating remotely [13].

ME sensors have emerged as particularly promising transducers in this context [14–19]. They are typically composed of amorphous ferromagnetic ribbons that can be mechanically excited by AC magnetic fields, allowing remote frequency reading without the need for wires or batteries. Their sensing principle relies on the high sensitivity of the resonance frequency to surface mechanical stress, which can be enhanced through functionalization with hygroscopic materials. Previous successful approaches include metal oxides ($Al_2O_3$ [20] $TiO_2$ [21,22]), cellulose nanofiber composites (CNFs-PVA) [23], and metal-organic frameworks (MOFs) [24]. In this work, we explore for the first time the conducting polymer PEDOT:PSS as a novel functional layer for ME-based humidity sensors. PEDOT:PSS combines flexibility with high humidity sensitivity [25], as its hydrophilic PSS component absorbs water. The charged ($-SO_3^-$) and neutral ($-SO_3H$) sulfonic groups interact with water molecules, leading to polymer swelling [26]. This hygroscopic response generates a measurable mass change, enabling wireless ME resonance sensing. The proposed approach provides a pathway for low-cost, highly sensitive humidity sensors with remote monitoring capabilities, leveraging controlled shifts in ME resonance frequency induced by PEDOT:PSS-humidity interactions.

## 2. Experimental Section

### *2.1. Preparation and characterization of ME humidity sensors*

The PEDOT:PSS films were fabricated on ribbon-shaped resonator platforms with dimensions of 2 mm × 10 mm × 28 µm, made from the METGLAS 2826MB3 alloy ($Fe_{40}Ni_{38}Mo_4B_{18}$). The substrates were previously cleaned with acetone, isopropyl alcohol, and deionized water to remove any organic contamination. After that, PEDOT:PSS polymer films were obtained by drop-casting, using a PEDOT:PSS (1:2.5) solution (weight ratio, concentration acquired from the manufacturer). It is worth noting that varying this ratio could tailor the film's electrical and viscoelastic properties, presenting an avenue for future sensor optimization. The film thicknesses reported in this work were directly measured using an optical profilometer. During deposition, we varied the dispensed volume per unit area (surface concentration) of the precursor solution 0.1, 0.25, 0.5, 0.75, and 1.25 µL/mm². These surface concentrations resulted in measured thicknesses of approximately 2 µm, 5 µm, 10 µm, 15 µm, and 25 µm, respectively. The samples were then annealed for 5 minutes at 60 °C after complete drying (see in Fig. 1). This annealing step was performed to promote better adhesion to the METGLAS substrate, and enhance the structural stability properties of the PEDOT:PSS coating.

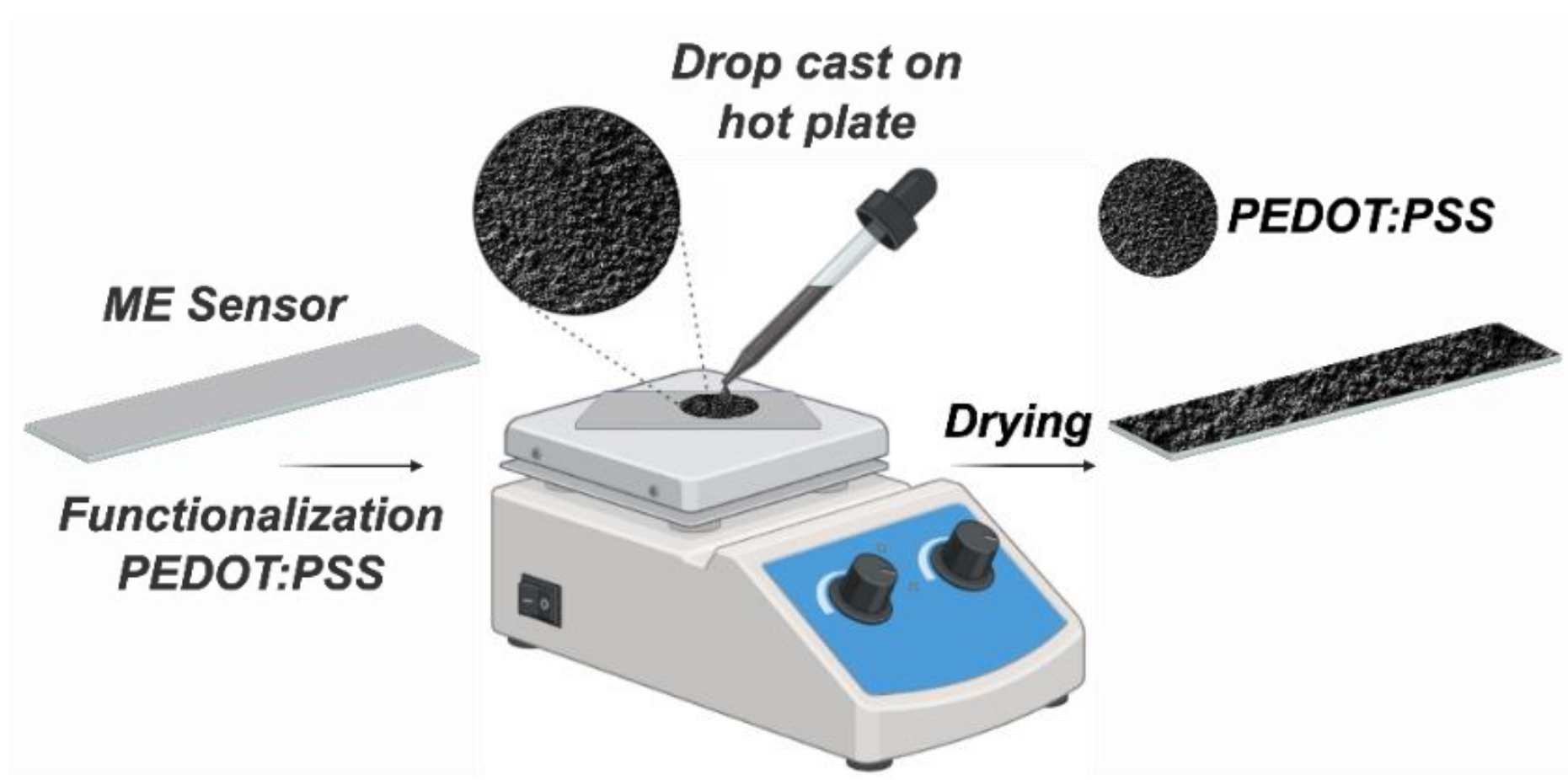

**Figure 1** - Schematic representation of the functionalization process of ME sensors using PEDOT:PSS polymer solution by drop-casting technique.

Comprehensive characterization of PEDOT:PSS films on the ME sensor surface was performed using Scanning Electron Microscopy (SEM), Energy-Dispersive X-ray Spectroscopy (EDX), Atomic Force Microscopy (AFM), Kelvin Probe Force Microscopy (KPFM), micro-Raman spectroscopy, and X-ray Diffraction (XRD) to evaluate morphology, composition, structure, and humidity response. SEM and EDX analyses were conducted using a JEOL-JSM-6010LA microscope at 15 kV and 7 kV, respectively. KPFM and AFM measurements were performed with a FlexAFM Nanosurf system in a hermetic chamber with controlled RH, dry $N_2$ flow was used to reduce the humidity level, allowing subsequent adjustment to the desired RH conditions. A conductive ElectricMulti75-G probe (Budget Sensors), coated with 5 nm Cr and 25 nm Pt, with a resonance frequency of 75 kHz and a force constant of 3 N/m, was used with a 1 V, 17 kHz modulation signal to map the contact potential distribution. Raman measurements were carried out using a Renishaw InVia system with a 514.5 nm laser, 50× objective, and power below 1 mW. X-ray Diffraction (XRD) analysis was performed using a Bruker D8 DISCOVER diffractometer with Cu $K\alpha_1$ radiation ($\lambda$ = 1.5406 Å), a Göbel mirror, and a Ge (220) monochromator to ensure a monochromatic and parallel incident beam.

*2.2 Principle of detection and resonance frequency measurements of ME humidity sensor*

The schematic in Fig. 2 depicts the experimental setup used to measure the resonance frequency of ME sensors subjected to variations in RH.

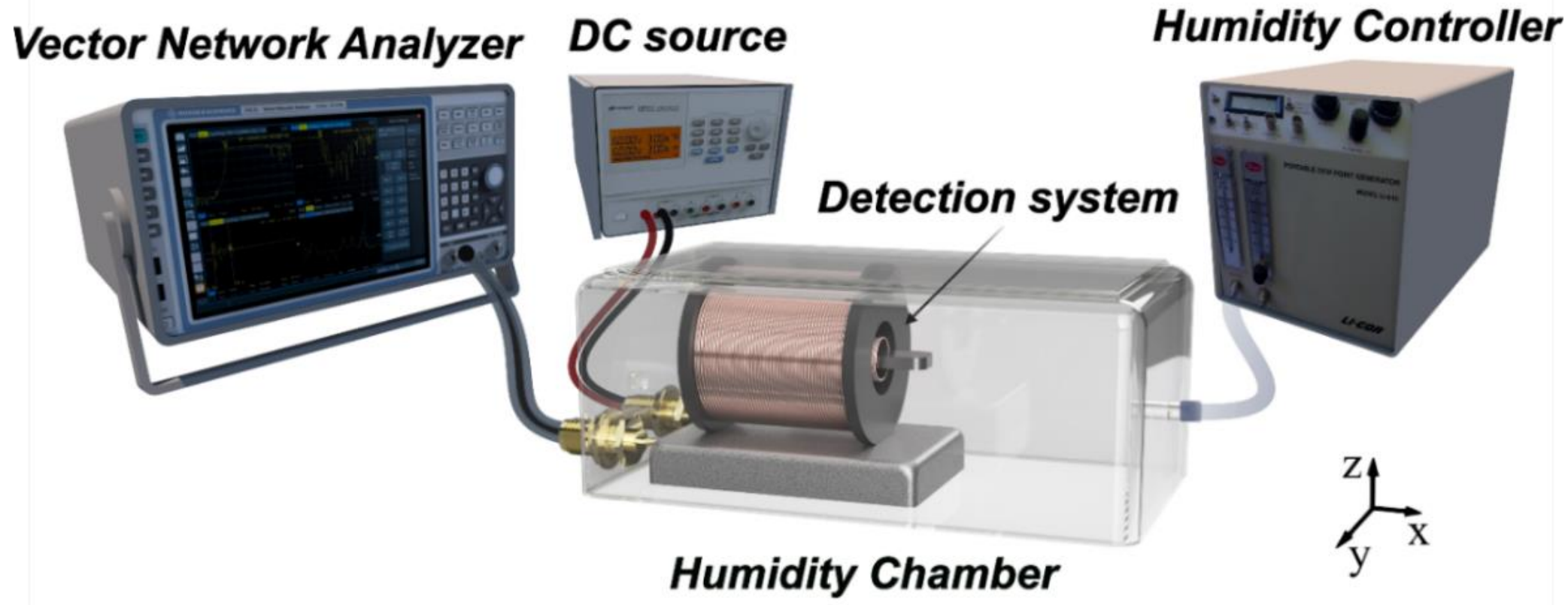

**Figure 2 -** Schematic of the experimental setup for measuring the resonance frequency of ME sensors as a function of RH using a VNA. A sealed humidity control chamber houses the sensor, pickup coil, and bias coil, and is connected to an RH controller.

A plane-wave-type perturbation is applied to the ME resonator through the application of magnetic fields $H = H_{DC} + H_{RF}(\omega)$ along the x-direction (see Fig. 2), causing the device to oscillate with angular frequency $\omega$, thereby reaching the resonance regime. Assuming that the magnetic fields are uniform along the x-axis (see Fig.2), the displacement vectors become independent of y, allowing the derivation of a one-dimensional wave equation for elastic solids [27]:

$$\frac{\partial^2 u_x}{\partial t^2} = \frac{E}{\rho_r(1-\nu^2)}\frac{\partial^2 u_x}{\partial x^2}, \qquad (1)$$

where $E(H)$ is the effective Young's modulus, modulated by the ΔE effect, $\rho_r$ is the density of the ME resonator, $u_x$ is a function describing the mechanical displacement of the system (resonator/film) in the x-direction over time, and $\nu$ is the Poisson's ratio. For a purely elastic and thin mass layer, an increase in mass on the sensor surface causes a downward shift $\Delta f$ in the resonance frequency, described by the Sauerbrey equation [28]:

$$\Delta f = -\frac{\rho h}{2\rho_r d} f_0, \qquad (2)$$

where $\rho$ and $h$ are, respectively, the density and thickness of the deposited mass layer, and $d$ and $f_0$ are the thickness and the fundamental resonance frequency of the uncoated ME sensor, respectively. However, this model serves only as a first approximation, as the PEDOT:PSS film exhibits significant viscoelasticity.

Assuming that the PEDOT:PSS film on the ME sensor behaves as a viscoelastic medium, the contact between the resonator and this layer leads to energy dissipation through shear, since the coupling occurs via tangential shear at the interface. According to Lang (2009), the Young's modulus of PEDOT:PSS decreases significantly with increasing RH [29], indicating a reduction in stiffness and mechanical cohesion due to hydration of the PSS-rich layer. Within the framework of the Kelvin-Voigt viscoelastic model, the key parameter for shear-dominated interactions is the complex shear modulus, $G^* = G' + j\omega\eta$, where $G' = E/2(1+\nu)$ is the storage modulus and $G'' = \omega\eta$ corresponds to the loss modulus, representing energy dissipation [30].

The displacement field $\xi_z(t)$ within the viscoelastic film, in response to the oscillating resonator surface with displacement $u_x(t)$, can be described by [31]:

$$\xi_z(t) = u_x \frac{cos\big(\gamma(h-z)\big)}{cos(\gamma h)} e^{j\omega t}, \qquad (3)$$

where $\gamma = \omega\sqrt{\rho/G^*}$ is the complex wave number, which governs the propagation and decay of the shear wave within the viscoelastic layer, $h$ is the film thickness, ρ is its density, and the boundary condition at the free surface is $\xi_0(t) = u_x(t)$ (coupling) and $(\partial\xi_z/\partial z)_{z=\mathrm{h}} = 0$. The shear stress $\sigma_{xz}$ at the interface ($z = 0$) is given by:

$$\sigma_{x0} = G^* \left(\frac{\partial \xi_z}{\partial z}\right)_{z=0} = G^* \gamma tan(\gamma h) u_x \, e^{j\omega t}. \qquad (4)$$

Equation (4) shows that the shear stress, and consequently the energy dissipation, depends on the complex shear modulus $G^*$. The hydration of the PSS-rich layer, which causes a strong decrease in the Young's modulus $E$ [29], softens the film (reducing $G'$) and enhances its dissipative character (affecting $G''$). This leads to increased damping. This leads to increased damping, as quantified by the loss tangent $tan(G''/G')$ [30]. Thus, in addition to the primary effect associated with the deposition of water vapor mass on the surface of the ME sensor, resulting from moisture absorption by PEDOT:PSS (equation 2), the variation of the complex shear modulus due to hydration also contributes significantly to the resonance frequency shift, since it modifies the mechanical properties of the film and alters the propagation velocity of acoustic waves in the resonator. These viscoelastic effects, more pronounced at high humidity levels, lead to greater signal damping. The combination of mass coupling and viscoelastic damping at hydrated film intensifies the sensor's dynamic response, increasing its sensitivity to humidity and making the device more efficient for RH monitoring applications.

The resonance frequency measurements carried out in our ME sensors were performed using a vector network analyzer (R&S ZNLE18, Rohde & Schwarz, see Fig. 2) operating in $S_{11}$ mode and connected to a double-layer excitation/detection solenoid coil (200 turns, 4 mm inner diameter, and 22 mm length), which generates the RF excitation signal to perform a frequency sweep and monitors the reflected signal. The DC magnetic induction field was applied using a second solenoid coil (600 turns, 23 mm inner diameter, and 50 mm length, made with 0.40 mm copper wire and having an overall electric resistance of 10.7 ohms), concentric with the excitation/detection coil and powered by a Keysight U8031A DC source. Both coils were placed inside a closed acrylic chamber, where the atmosphere consisted of ambient air. The resonance frequency was determined from the $S_{11}$ parameter. An alternative measurement system to the VNA, based on low-cost microcontrollers, can be used for the determination of the ME resonance frequency [14].

The RH inside the chamber (with dimensions of $12 \times 7 \times 7$ cm$^3$) was controlled using a LI-610 dew point generator (LI-COR) with a resolution of 0.1% RH. The instrument uses ambient air as input and adjusts it to the desired RH level before delivering it into the chamber. The desired RH is generated by controlling the dew point of the incoming air stream. A low air flow rate was maintained throughout the measurements to ensure response stability. Measurements were performed by sweeping the RH from 20% to 95%, while the $S_{11}$ parameter was continuously monitored using the VNA to obtain the response curve of the ME sensor. For the determination of response and recovery times, nitrogen ($N_2$) was used to purge the chamber and establish low-humidity conditions prior to the introduction of air at the desired RH level. The response and recovery times were extracted from the transient frequency-time curves during controlled RH transitions using the 90% criterion. Measurements were performed with the sensor mounted on a holder and inserted into the excitation/detection coil. All experiments were conducted at a fixed room temperature of 22 °C.

## 3. Results and Discussion

The composite polymer PEDOT:PSS is a mixture of poly(3,4-ethylenedioxythiophene), with monomer of $C_6H_4O_2S$ composition, and a polystyrene sulfonate with $C_8H_8O_3S$ (or $C_8H_7O_3S$ depending on the sulfonic termination) composition [32]. The EDX spectrum in Fig. 3a shows only the expected elements from PEDOT:PSS and from the ME substrate, with no detectable heavy contamination, and SEM imaging (inset of Fig. 3a) confirms the uniform coverage of the film. Raman spectroscopy (Fig. 3b) displays the characteristic vibrational bands of PEDOT:PSS in agreement with the literature, indicating that the polymeric configuration is preserved after deposition, with PSS-associated vibrational modes. Additional information on the normal modes is provided in the Supplementary Information.

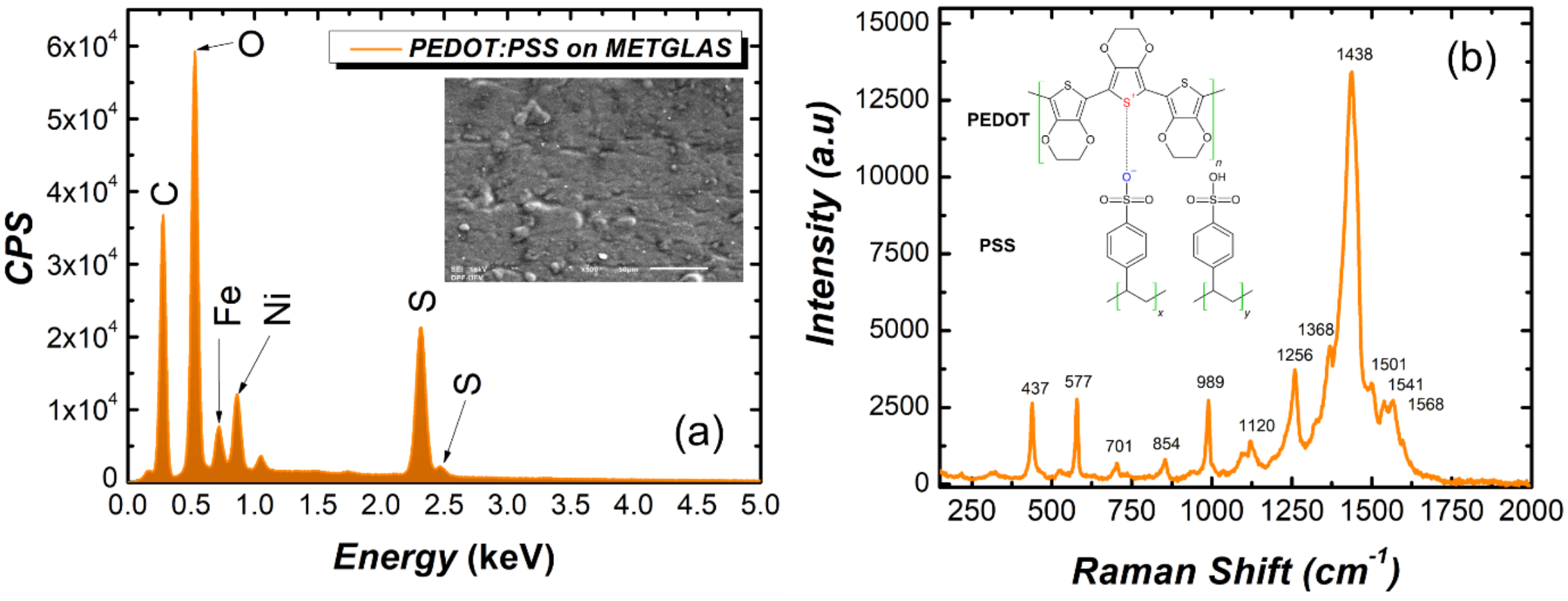

**Figure 3** - (a) EDX spectrum and SEM inset image showing the elemental composition and surface morphology of the PEDOT:PSS film deposited on the magnetoelastic (ME) resonator. (b) Raman spectrum of the PEDOT:PSS film, highlighting the characteristic vibrational modes of the polymer.

To investigate the role of RH on the structural and electronic surface characteristics of PEDOT:PSS films, AFM and KPFM measurements were performed under controlled ambient conditions. Fig. 4a and Fig. 4b show the AFM and KPFM images obtained at low RH (5 ± 2)%, where a rougher morphology with finer grains is observed, along with positive contact potential difference (VCPD) values, indicating a lower work function relative to the Au-coated conductive tip. Under these dry conditions, the limited presence of water molecules restricts polymer chain mobility, hindering surface reorganization and resulting in finer grains with increased roughness, as confirmed by the AFM topography (Fig. 4a). These observations are consistent with previous reports of reduced molecular rearrangement under anhydrous conditions [1,2]. At higher humidity (RH > 60%), as shown in Fig. 4c and Fig. 4d, the surface becomes smoother with larger domains, while the surface potential shifts to negative values, indicating an increase in the work function. This behavior is attributed to water absorption by the hydrophilic PSS chains, which enhances polymer mobility and promotes structural rearrangement [1,33]. These results indicate a clear correlation between humidity and both morphological and electronic surface properties, with increasing RH leading to a more uniform topography and a significant shift in surface potential. A detailed discussion of these effects is provided in the Supplementary Information.

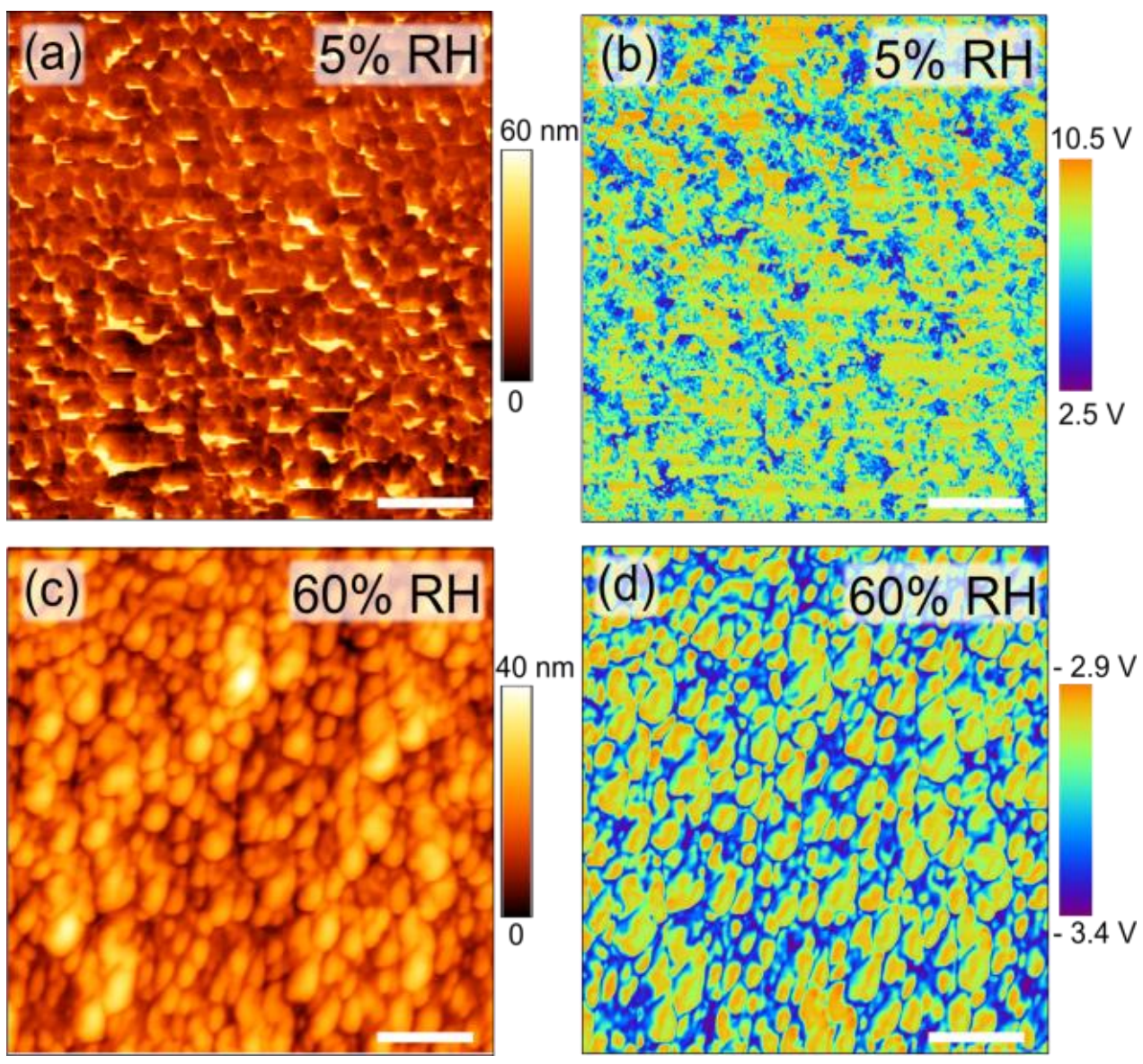

**Figure 4** - AFM images (a, c) and corresponding KPFM maps (b, d) of PEDOT:PSS films at 5% RH (a, b) and 60% RH (c, d). Scale bars: 200 nm.

To lamellar structure to environmental humidity, XRD patterns were recorded under controlled RH conditions, ranging from 20% to 95% (Fig 5a and Fig. 4b). Fig. 4b, a progressive shift of the (100) reflection toward lower diffraction angles was observed with increasing humidity, indicating an expansion of the lamellar spacing. This expansion becomes particularly pronounced at high RH values (≥ 60%), where the PSS-rich domains absorb water and undergo significant swelling. Water molecules disrupt the initial packing density, leading to a larger interlayer separation in consonance with AFM/KPFM results. The evolution of d-spacing values extracted from the (100) diffraction peak versus RH (Fig. 5b, inset). The (100) lamellar spacing increases progressively from 23.5 Å under dry condition to 23.9 Å at 60 % RH, and reaches 24.2 Å at 95 % RH. This substantial increase in lamellar spacing highlights the high affinity of PSS for water and the resulting volumetric response of the nanostructure. This behavior is similar to that observed in fiber-like PEDOT:PSS structures, reflecting the intrinsic structural properties of the polymer [34]. Interestingly, the (020) reflection remains relatively unaffected by humidity variations, suggesting that the π-π stacking of PEDOT chains remains stable even as the lamellar framework undergoes expansion. This indicates that the crystalline regions responsible for electrical transport may retain structural integrity despite environmental fluctuations [34]. Additional information regarding the results of the XRD is provided in the Supplementary Information.

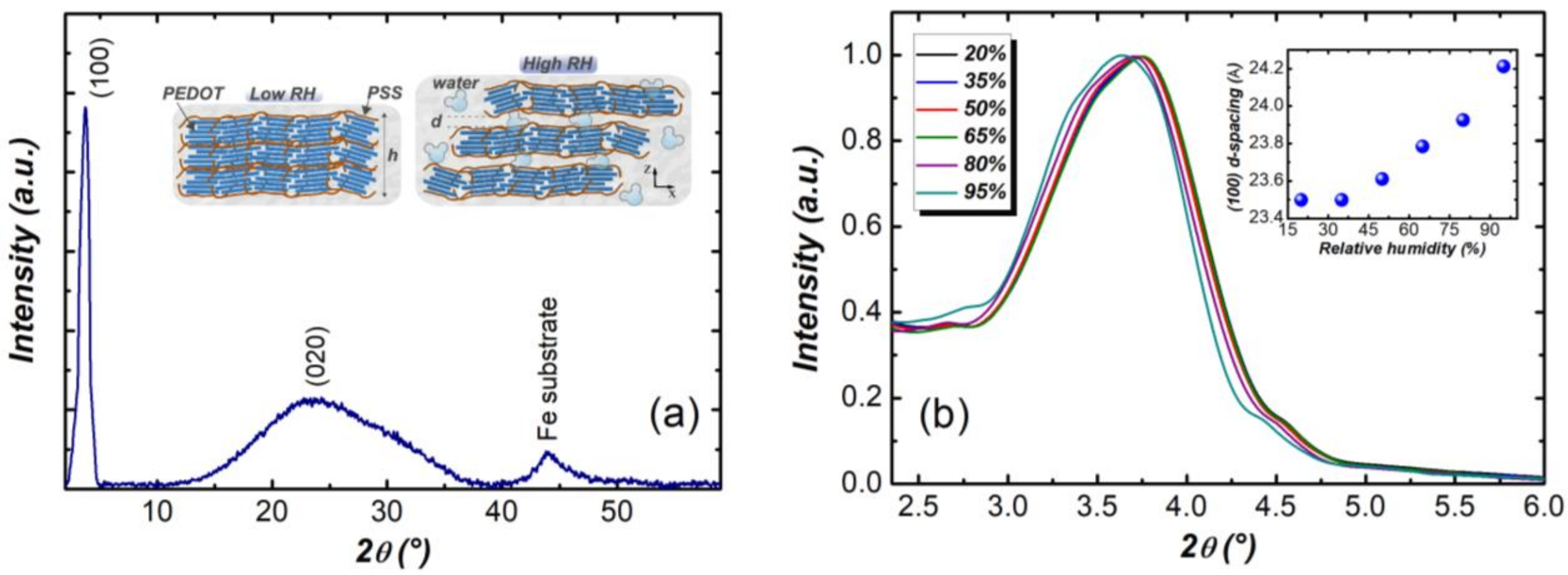

**Figure 5** - (a) XRD pattern of PEDOT:PSS films on METGLAS substrates, highlighting reflections related to the polymer structure. (b) Zoom of the (100) peak region, showing a shift toward lower angles with increasing RH, indicating lamellar expansion. *Inset*: interplanar distance $d_{100}$ as a function of RH, showing a nearly linear increase; marker size is comparable to the error bars.

Figure 6 presents complete spectra in the vicinity of the sensor resonance frequency (Fig. 6a) and the quality factor, damping and minimum amplitude at resonance (inset) (Fig. 6b) for the ME sensor functionalized with 0.5 μL/mm² (10 μm) of PEDOT:PSS, subjected to different RH levels. A systematic decrease in resonance frequency is observed with increasing humidity, which is consistent with the adsorption of water molecules by the PEDOT:PSS film. The hydration of the polymer film increases the resonator's surface mass, as described by the Sauerbrey equation, and significantly alters its viscoelastic properties, an effect quantified by the shear stress at the resonator-film interface (Eq. 4). Hence, the combined effect of changing RH results in both frequency shift and more pronounced signal damping. In the case of PEDOT:PSS, higher RH levels promote hydration of the PSS phase, leading to partial disruption of hydrogen bonds and enhanced mobility of the polymer chains. As a consequence, the film exhibits an increased effective viscosity, which consequently intensifies mechanical energy dissipation when the system is under oscillation. This behavior aligns with the decrease in Young's modulus reported by Lang (2009) for hydrated films, previously reported to become more plastic and less rigid [29].

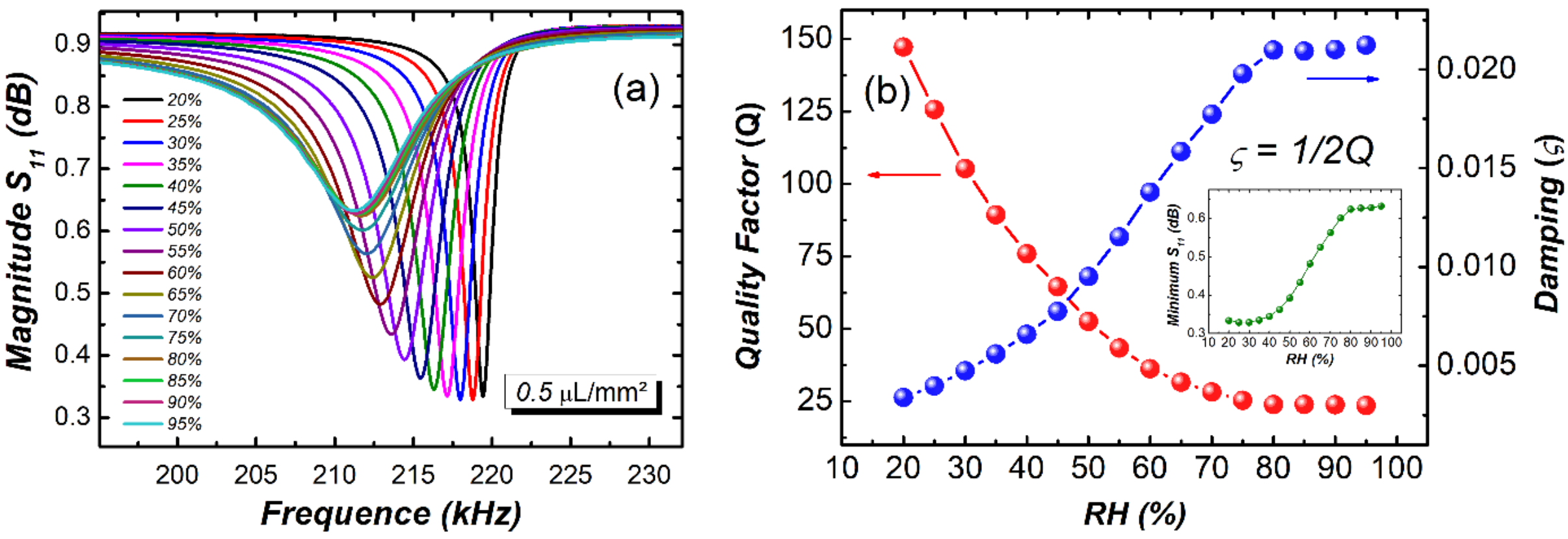

**Figure 6** - (a) The change in resonance frequency and, (b) quality factor, damping and minimum amplitude at resonance (*inset*) for the sensor with a 0.5 µL/mm² coating under different RH values.

The greater deformability allows out-of-phase motion between polymer domains, broadening the resonance curve and increasing damping, as evidenced in Fig. 6b, where a decrease in the resonance quality factor is accompanied by an increase in the damping coefficient. This highlights the energy dissipation mechanism that, in the case of the sensor, contributes to the enhanced sensitivity. The correlation with the X-ray diffraction results (Fig. 6b), which point to an increase in the interplanar spacing (d-100) with humidity, confirms the swelling of the PSS phase and the resulting structural reorganization of the film. Quantitatively, the full width at half maximum (FWHM) of the resonance peak increased from approximately 1490 Hz at 20% RH to 8970 Hz at 95% RH, directly reflecting the rise in damping. These findings reinforce the interpretation that the sensor's response is governed by both mass loading and humidity-induced viscoelastic effects.

Figure 7a shows the frequency response of ME sensors with different PEDOT:PSS coating thicknesses, where intermediate films provide smoother and more stable behavior. The sensor with a 0.5 µL/mm² coating (10 µm thickness) was selected as the optimal configuration, and its response is detailed in Fig. 7b. The nonlinear behavior indicates competing effects, such as mass loading saturation and increased viscoelastic damping at higher RH levels. Figure 7c present the resolution, calculated from the frequency noise ($\sigma$ = 13.8 Hz) [35]. In the 20 - 70% RH range, the sensor exhibits a resolution of approximately 0.1 %RH and an accuracy characterized by a mean absolute error of 1.2 %RH. In the higher humidity range (70 - 95% RH), the resolution degrades to about 0.5 %RH, while the accuracy is 1.1 %RH.

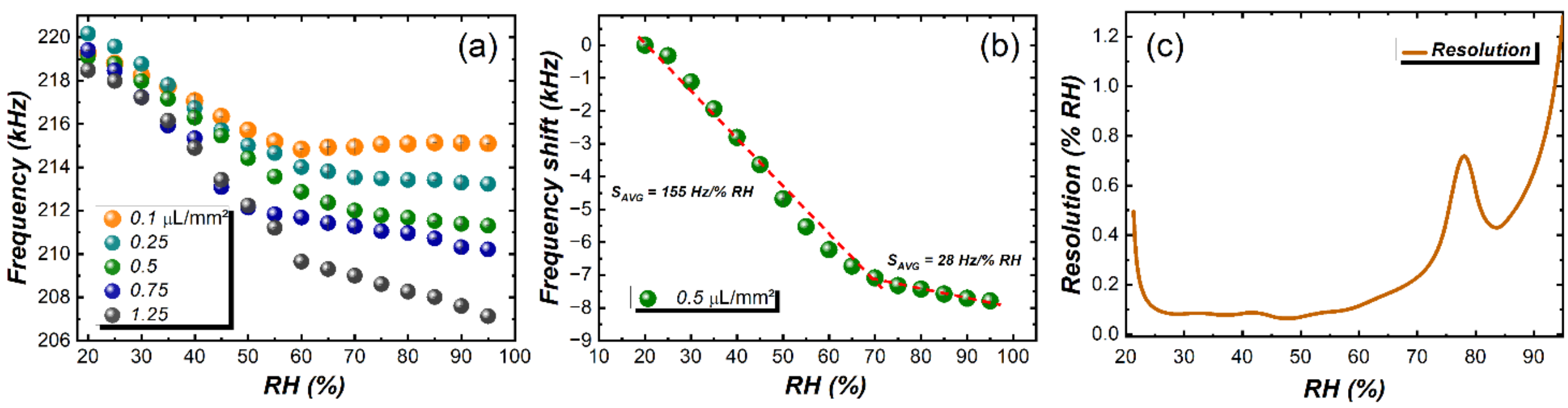

**Figure 7 -** (a) Resonance frequency shift as a function of RH for different PEDOT:PSS coatings analyzed (0.1, 0.25, 0.5, 0.75, and 1.25 μL/mm²). (b) Shift resonance frequency as a function of RH for a 0.5 μL/mm² coating (10 μm thickness), (c) Resolution for the sensor with coating 0.5 μL/mm², selected as the reference.

The sensitivity $S$ for the sensor shown in Fig. 7b is particularly high in the range of 20 - 70% RH ($S$ = 155 Hz/% RH), but decreases between 70 - 95% RH ($S$ = 28 Hz/% RH), obtained through a linear fit of the data in the graph in Fig. 7b in these operating ranges (defined as the slope of the linear region of $f$ vs RH). The limit of detection (LOD) was estimated as LOD = $3\sigma/S$, yielding approximately 0.3% RH in the 20 - 70% RH range and 1.5% RH in the 70 - 95% RH range. The lowest experimentally tested relative humidity was 5% RH. For the coating of 1.25 μL/mm², an even higher sensitivity is achieved in the 20 - 70% RH range ($S$ = 223 Hz/%RH), which also decreases in the 70 - 95% RH range ($S$ = 71 Hz/%RH). This sensitivity is particularly higher than those found in recent studies with ME humidity sensors [22–24]. The sensitivities and resolutions for the remaining sensors were calculated and incorporated into the Supplementary Information. Table 1 compares the proposed ME sensor with other wireless detection devices using different humidity-sensitive coatings. Compared to QCM-based sensors in the literature, it shows comparable or higher sensitivity and the advantage of contactless operation [36]. For example, a PEDOT-based QCM humidity sensor showed a sensitivity of around 33 Hz/%RH at the 40 - 60% RH range [37].

**Table 1 -** Performance comparison between the ME humidity sensor. * indicates the optimized sensing layer thickness used in this work.

| Technique | Ref. | Sensing material | Range (%) | Sensitivity (Hz/%RH) | τ90 up (s) | τ90 down (s) |
|---|---|---|---|---|---|---|
| ME | **This work** | PEDOT:PSS(0.5μm)/Metglas 2826MB3* | 20 - 70 | 155 | 22 | 11 |
| | | PEDOT:PSS(1.25μm)/Metglas 2826MB3 | 20 - 60 | 223 | 22 | 11 |
| ME | Liu et al. 2024 [23] | CNFs-PVA/MetglasTM 2826MB | 54 - 85 | 23.8 | ~ 900 | none |
| ME | Atalay et al. 2020 [22] | TiO2-NT film on Metglas 2826MB | 5 - 95 | 35.3 | 40 | 60 |
| ME | Grimes et al. 2000 [21] | TiO2-solgel on Metglas 2826MB | 2 - 70 | ~1.5 | none | none |
| QCM | Jebnouni et al. 2026 [38] | LiCl-loaded ZnO | 0 - 98 | 48.1 | 11 | 5 |
| LC Resonant Circuit | Meng et al. 2025 [39] | $In_2O_3$/GO | 30 - 60 | $1.0 \times 10^6$ | 25 | 80 |
| LC Resonant Circuit | Zhou et al. 2026 [40] | $Ti_3C_2T_x$ MXene/$WS_2$ | 10 - 60 | $1.7 \times 10^6$ | 17 | 25 |
| LC Resonant Circuit | Hou et al. 2026 [41] | MoS2/CS | 40 - 60 | $0.06 \times 10^6$ | 70 | 120 |

Additional changes in viscosity, Young's modulus, and interplanar spacing under humid conditions may not be fully reversible over short time scales, leading to a small memory effect in the sensor response. As shown in Fig. 8a, sensors S1-S3 exhibit a stable and repeatable resonance frequency response during successive RH cycles, indicating good repeatability and reproducibility in both fabrication and sensing behavior. Fig. 8b shows a zoom of one cycle of sensor S3, where the response and recovery times can be seen in more detail. The response and recovery times extracted from the humidity cycles for sensors S1-S3 are presented in Fig. 8c. The average values were 22 ± 1 s for humidification (up) and 11 ± 1 s for dehumidification (down). This difference indicates that the desorption process is kinetically faster than the absorption and diffusion of water molecules into the polymer matrix, which is consistent with the rapid structural relaxation of the film during drying. Hysteresis was estimated from the cyclic response of sensor S3, yielding an average value of 0.5 ± 0.2 % over five cycles. After two months of storage under ambient atmospheric conditions, the sensor was measured again to evaluate its long-term stability. The response and recovery times of 22 ± 6 s for humidification (up) and 11 ± 2 s for dehumidification (down) were obtained (see Fig. S1 in the Supplementary Information), together with a reduction in the sensitivity from the initial post-fabrication value of $S$ = 109 Hz/%RH to $S$ = 84 Hz/%RH. This analysis indicates that there was no significant change in the dynamic response of the sensor, since the response and recovery times remained similar. The decrease in sensitivity may be related to a change in the mechanical coupling of the polymer with the resonator, which affects the transduction capability.

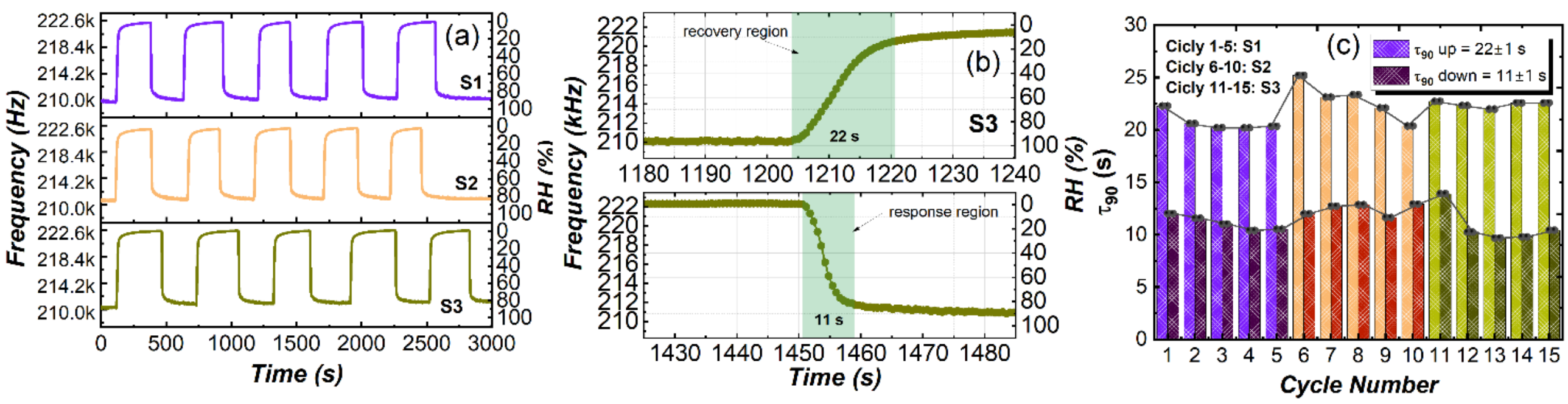

**Figure 8 -** (a) Resonance frequency response of sensors S1-S3 (0,5 μL/mm²) under repeated RH cycling. (b) Enlarged view of one RH cycle for sensor S3, highlighting the response and recovery times. (c) Response and recovery times ($\tau_{90}$) for 15 humidity cycles obtained from the sensors shown in (a), with mean values of $\tau_{90}$ up = 22 ± 1 s and $\tau_{90}$ down = 11 ± 1 s.

Although this study demonstrates high performance in ambient air, the sensor's response in complex environments warrants discussion. The detection mechanism relies on water absorption by the PSS phase, conferring inherent selectivity to humidity [33]. Since humidity is a critical and ubiquitous interference factor in practical gas sensing, the sensor's response inherently reflects realistic operational conditions [42]. However, temperature variations can affect the response by altering the water absorption equilibrium and the viscoelastic properties of the polymer. For precise field applications, future work should include temperature compensation and explicit selectivity tests against specific volatile organic compounds (VOCs). Monitoring RH in sealed environments, such as food packaging and silos, where the atmosphere is dominated by $N_2$ and $CO_2$ [12], highlights the application potential of the sensor. This scenario motivates the development of wireless and passive sensing platforms for remote operation. Ongoing work focuses on optimizing a low-cost experimental detection setup based on microcontrollers, with the goal of applying the developed sensors to measurements inside food packaging. Specifically, the sensor placed inside the packaging will be wirelessly interrogated by means of an external pickup coil, allowing non-invasive monitoring.

## 4. Conclusion

This work demonstrated the feasibility of employing PEDOT:PSS as a functional coating for magnetoelastic resonators, enabling the development of wireless humidity sensors with high sensitivity and stability. The interplay between water adsorption, structural swelling of the PSS domains, and viscoelastic damping governs the resonance response, leading to high sensitivity and improved humidity resolution. Structural and surface analyses confirmed the integrity of the polymer and revealed humidity-driven swelling of PSS domains, accompanied by increased chain mobility and lamellar spacing. These results highlight PEDOT:PSS as a promising coating material for magnetoelastic humidity sensing, demonstrating its effectiveness in enabling low-cost, wireless sensors with high sensitivity and resolution suitable for monitoring RH in sealed or hard-to-access environments.

**Declaration of Competing Interest**

The authors declare no known financial interests that may be perceived as a conflict of interest in relation to this work.

**Acknowledgment**

This research is supported by the Conselho Nacional de Desenvolvimento Científico e Tecnológico (CNPq), Coordenação de Aperfeiçoamento de Pessoal de Nível Superior (CAPES), Financiadora de Estudos e Projetos (FINEP), Fundação de Amparo à Pesquisa do Estado de Minas Gerais (FAPEMIG) - Rede de Pesquisa em Materiais 2D and Rede de Nanomagnetismo, INCT Nanocarbono e Materiais 2D, CNPq 408649/2024-0, and INCT of Spintronics and Advanced Magnetic Nanostructures (INCT-SpinNanoMag), CNPq 406836/2022-1.

**Supporting information**

Supplementary data associated with this article can be found in the online version.

**Data availability**

Data will be made available on request.

**Supplementary Information**

**PEDOT:PSS-Coated Magnetoelastic Sensors for Highly Sensitive Wireless Humidity Sensing**

**A.** ***Regarding the characterization and performance evaluation of the PEDOT:PSS film***

Raman spectroscopy was carried out to investigate possible modifications in the PEDOT:PSS structure induced by the deposition process. The set of peaks detected by this method, shown in Fig. 3b, is related to vibrational modes of PEDOT:PSS and the peak positions are strongly dependent on the laser excitation energy [1]. The retrieved spectrum exhibits excellent agreement with previous studies [1–8], indicating that the polymeric configuration is preserved after the functionalization process. Details of the Raman spectrum of pristine PEDOT:PSS can be understood by the dominance of vibrational modes of carbon and oxygen atoms. The main peaks related to this composite polymer structure are listed as follows: at 437 $cm^{-1}$, $SO_2$ bond bending can be observed; at 577 $cm^{-1}$ and 989 $cm^{-1}$, oxyethylene ring deformations appear. The symmetric C−S−C deformation is related to the peak at 701 $cm^{-1}$, and C−O−C deformation give rise to the peak observed at 1093 $cm^{-1}$. Carbon bond vibrations are observed at 1256 $cm^{-1}$ ($C_\alpha$–$C_\alpha$ inter-ring), 1368 $cm^{-1}$ ($C_\beta$–$C_\beta$), 1421 $cm^{-1}$ (symmetric $C_\alpha$=$C_\beta$(−O)). At 1532 $cm^{-1}$ and 1563 $cm^{-1}$ one observes peaks that arise due to asymmetric deformations of the $C_\alpha$=$C_\beta$ bond [3]. The peaks related to the PSS part are ascribed to the 989 $cm^{-1}$, 1120 $cm^{-1}$, and 1570$cm^{-1}$ peak positions [8]. The narrow peak around 1444 $cm^{-1}$, also associated to the PSS, indicates a more homogeneous distribution of electronically conjugated domains and a regular structural organization [1].

To investigate the role of RH on the structural and electronic surface characteristics of PEDOT:PSS films, a series of measurements were carried out using Atomic Force Microscopy (AFM) and Kelvin Probe Force Microscopy (KPFM). These complementary techniques enabled the simultaneous characterization of topography and local surface potential under controlled ambient conditions. Fig. 4a and Fig. 4b show AFM and KPFM images, respectively, obtained under low RH conditions of (5 ± 2) %. In this regime, the KPFM measurements revealed positive values of contact potential difference ($V_{CPD}$), indicating that the work function of the PEDOT:PSS surface is lower than that of the Au-coated conductive AFM tip. This behavior is associated with a relatively unstructured or minimally rearranged surface, where the conductive PEDOT phase remains relatively more exposed, and the insulating PSS component is (in comparison with PEDOT) more embedded within the bulk of the film with. Under such dry conditions, the limited presence of water molecules in the environment restricts the mobility of polymer chains, which hinders substantial surface reorganization. The consequence is a finer grain structure with more pronounced surface roughness, as confirmed by the AFM topographic analysis (Fig. 4a). Such observations are consistent with previous studies reporting reduced molecular rearrangement and structural relaxation under anhydrous conditions [9,10].

In contrast, Fig. 4c and Fig. 4d present the AFM and KPFM data acquired under high-humidity conditions (RH > 60%). In this regime, the surface potential shifts to negative values, indicating an increase in the sample's work function. This phenomenon is attributed to the absorption of water molecules by the hydrophilic PSS chains, which are known to have a strong affinity for moisture. The presence of absorbed water acts as a plasticizer, significantly enhancing the mobility of the polymer [11–13]. As a consequence, the PSS phase can migrate towards the surface, while PEDOT chains tend to aggregate into larger domains. This enhanced chain mobility under humid conditions facilitates phase separation and the formation of more coalesced PEDOT-rich regions, leading to a smoother surface morphology and larger grain size, as observed in the AFM topography (Fig. 4c). Moreover, the absorbed water may contribute to the formation of an electrical double layer at the polymer-air interface. This effect, driven by the accumulation of mobile ionic species such as hydrated anions, has been previously shown to influence the surface work function and electrostatic potential [14].

To further explore the influence of RH on the structural organization of PEDOT:PSS films, X-ray diffraction (XRD) measurements were carried out on samples deposited onto ME substrates. The diffraction patterns, presented in Fig. 5a, reveal two characteristic features that reflect the internal nanostructure of the PEDOT:PSS composite. A sharp diffraction peak centered at $2\theta \approx 26.1°$ corresponds to the (020) reflection, which is associated with the π-π stacking of the PEDOT chains [15]. This reflection indicates the presence of a short-range ordered phase in which conjugated PEDOT backbones are arranged in a face-to-face configuration, promoting intermolecular charge transport. The intensity and position of this peak serve as indicators of the degree of π-stacking order, a key parameter influencing the electrical conductivity of the film.

At lower scattering angles ($2\theta < 5°$ at Fig. 5a), a sharp and intense diffraction peak is observed, which is attributed to the (100) reflection of PEDOT:PSS structure [15]. The relatively narrow width of the diffraction peak of our measurements with respect to previous works [16,17] points out to domain sizes of 8.5 nm (using Scherrer's equation in the first diffraction peak) [18,19]. This size suggests a relatively high degree of structural coherence along the stacking direction, where usual domain size retrieved in the cited works was found to be under 4 nm. Concomitantly, the enhanced intensity of the low-angle peak indicates a well-established long-range periodicity, pointing to a molecular registry at the scale of ten nanometers. Similar correlations between peak width, coherence length, and stacking order have also been reported for other polymeric and layered systems [20–22]. It should be noticed in our case that the peak intensity can also be influenced by factors such as crystallinity and preferred orientation (texture) [18,19], both strongly influenced by the deposition parameters, which were optimized here for device performance and resulted in the retrieved molecular [19].

The high sensitivity of the PEDOT:PSS-based ME sensor is fundamentally rooted in the synergistic interplay of two mechanisms theoretically outlined of Methods: effective mass loading and pronounced viscoelastic damping. The hydrophilic PSS domains exhibit substantial water uptake, leading to a significant mass increase on the resonator surface, per the Sauerbrey equation (Eq. 2). Concurrently, water absorption plasticizes the polymer matrix, causing swelling of the PSS-rich regions (as confirmed by XRD) and a drastic reduction in the Young's modulus. This directly enhances the dissipative behavior, as governed by the complex shear modulus $G^*$. The ME resonator's exceptional sensitivity to these nanoscale changes both in added mass and in the humidity-tunable viscoelastic properties of the coating amplifies the resonance frequency shift. This dual mechanism, where viscoelastic damping complements mass loading during water uptake, contributes to the humidity sensing response of the device, surpassing the performance of many ME devices that rely primarily on mass-loading effects alone [23–26]. Importantly, the increase in lamellar spacing with RH exhibits an approximately linear trend up to ~30% RH, and this behavior was found to be reproducible and reversible upon humidity cycling. Such reversibility confirms that the structural changes are not permanent but result from dynamic water absorption and desorption processes, primarily localized within the PSS phase. After returning to lower humidity levels, the structure gradually recovers its original configuration. These results emphasize the dynamic nature of PEDOT:PSS nanostructure and its susceptibility to environmental moisture.

### *B. Sensor response characterization*

Regarding long-term stability, the measurements were repeated using sensor S3 after exactly two months of storage under ambient atmospheric conditions, and the results are presented in Fig. S1. Average response and recovery times of 22 ± 6 s for humidification (up) and 11 ± 2 s for dehumidification (down) were obtained, together with a reduction in the sensitivity to S = 85 Hz/% RH.

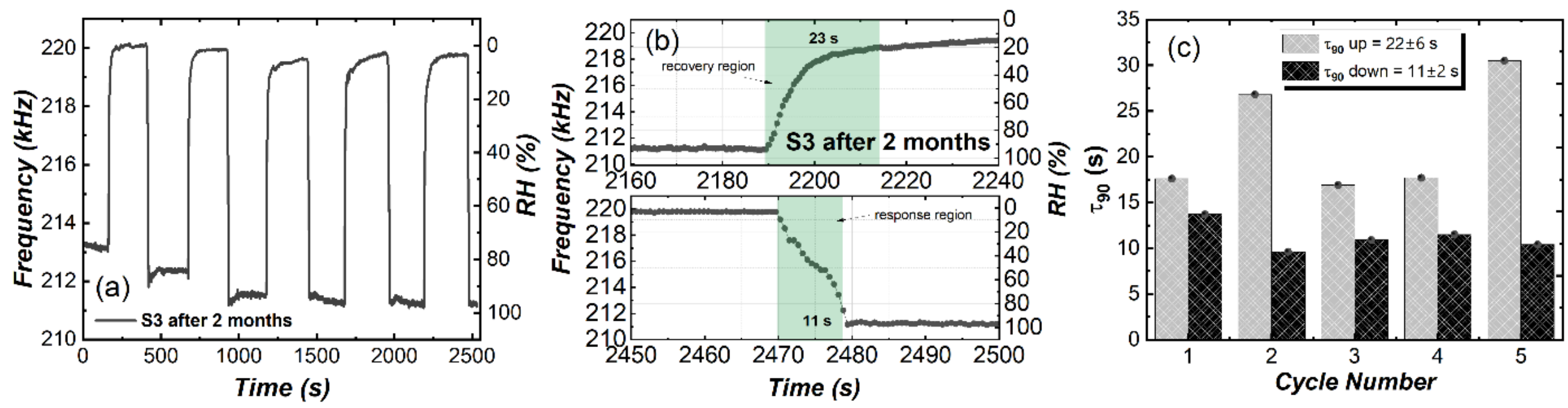

**Figure S1 -** (a) Resonance frequency response of sensor S3 after 2 months of storage. (b) Enlarged view of one RH cycle for sensor, highlighting the response and recovery times. (c) Response and recovery times ($\tau_{90}$) obtained from the sensor shown in (a), with mean values of $\tau_{90}$ up = 22 ± 6 s and $\tau_{90}$ down = 11 ± 2 s.